\documentclass[sigconf,nonacm]{acmart}
\AtBeginDocument{%
  }

\usepackage{todonotes}
\usepackage{tabularx}
\usepackage{todonotes}
\usepackage{multirow}
\usepackage{amsmath} 
\usepackage[utf8]{inputenc}

\begin{document}

\title{Closed-Loop Rhythmic Haptic Biofeedback via Smartwatch for Relaxation and Sleep Onset}


\author{Jueun Lee}
\email{jueun.lee@kit.edu}
\orcid{0000-0002-9862-9194}
\affiliation{%
  \institution{Karlsruhe Institute of Technology}
  \city{Karlsruhe}
  \country{Germany}
  }

\author{Dennis Moschina}
\email{moschina@teco.edu}
\orcid{0009-0003-9343-9150}
\affiliation{%
  \institution{Karlsruhe Institute of Technology}
  \city{Karlsruhe}
  \country{Germany}
  }

\author{Supraja Ramesh}
\email{supraja.ramesh@kit.edu}
\affiliation{%
  \institution{Karlsruhe Institute of Technology}
  \city{Karlsruhe}
  \country{Germany}
  }

\author{Tobias Röddiger}
\email{tobias.roeddiger@kit.edu}
\orcid{0000-0002-4718-9280}
\affiliation{%
  \institution{Karlsruhe Institute of Technology}
  \city{Karlsruhe}
  \country{Germany}
  }

\author{Kai Kunze}
\email{kai.kunze@keio.jp}
\orcid{0000-0003-2294-3774}
\affiliation{%
  \institution{Keio University}
  \country{Japan}
  }

\author{Michael Beigl}
\email{michael.beigl@kit.edu}
\orcid{0000-0001-5009-2327}
\affiliation{%
  \institution{Karlsruhe Institute of Technology}
  \city{Karlsruhe}
  \country{Germany}
  }

\renewcommand{\shortauthors}{Lee et al.}

\begin{abstract}
We investigate the use of musically structured, closed-loop vibration patterns as a passive biofeedback intervention for relaxation and sleep initiation. By encoding rhythmic meter structures into smartwatch vibrations and adapting their frequency to be slightly slower than the user’s real-time heart rate, our system aims to reduce arousal through tactile entrainment, offering a non-invasive alternative to auditory or open-loop approaches previously used in sleep and anxiety contexts.
In the first study (N=20), we compared five adaptive vibration rhythms for their effects on heart rate and subjective perceptions of relaxation in a resting context. In the second study (N=28), we evaluated the most promising pattern from Study 1 in a prolonged sleep initiation setting. Results showed increased parasympathetic activity and perceived relaxation during short-term stimulation, but no significant effects on sleep-related measures during the sleep onset phase.
This work contributes to the understanding of how wearable haptic feedback can support relaxation and sleep, offering design insights and identifying methodological considerations for effectively integrating haptic interaction into self-directed interventions.
\end{abstract}



\maketitle

\section{Introduction}
\label{sec:introduction}

Achieving restful sleep is essential for maintaining overall well-being, yet difficulties in falling asleep and attaining relaxation remain pervasive challenges \cite{chattu2018global, murawski2018systematic, american2010resolution}. Disruptions to natural sleep onset are often influenced by factors such as heightened psychological tension \cite{carskadon2004regulation}. Stress-related physiological responses can interfere with the body’s ability to smoothly transition into sleep \cite{martire2020stress, horvath2016effects} and are strongly associated with prolonged sleep latency and fragmented sleep \cite{sateia2000evaluation, pagel2001medications}. Promoting sleep for a healthy lifestyle involves recognizing it as a modifiable behavior that can be enhanced through supportive, non-invasive techniques \cite{morin2011oxford}. 

Wearable haptic interventions have recently emerged as a promising method for mitigating everyday stress with the widespread integration of vibration feedback in mobile technologies \cite{woodward2020beyond, vyas2023descriptive}. Unlike visual or auditory cues, haptic feedback can directly influence affective physiological states through non-cognitive pathways, enabling immediate emotional modulation with minimal cognitive effort \cite{hertenstein2009communication, knapp1978nonverbal, van2015social}, referred to as "passive intervention" \cite{zhao2023affective}. 
In particular, haptic biofeedback modulates physiological states by leveraging real-time bodily signal, enabling pre-conscious emotion regulation \cite{jain2020designing, dobrushina2024training, chua2024know}. Prior biofeedback researchers using wearable haptics has demonstrated effectiveness in implicit down-regulation of physiological arousal and enhancing physiological synchrony \cite{slovak2023designing, palumbo2017interpersonal}.

To further support relaxation at sleep onset, we draw on a modality long associated with calming physiological effects: music. Lullabies, in particular, exhibit features such as slow tempos, regular meters (often 2/4 or 3/4), legato articulation, and repetitive phrasing \cite{trehub1993maternal, unyk1992lullabies}, consistently identified in computational analyses of sleep-related music \cite{jespersen2022lullaby, scarratt2023audio}. These structures are  linked to reductions in arousal and heart rate, as well as modulation of stress-related biomarkers \cite{bernardi2006cardiovascular, khalfa2003effects, ooishi2017increase}, and are thought to promote entrainment with internal rhythms like respiration and cardiac cycles \cite{hesse2013musik, yamasato2019characteristics, fernandez2016influence}.

Building on these findings, we explore how musically informed haptic rhythm patterns, modulated through closed-loop biofeedback, can serve as a passive intervention to support relaxation during the transition to sleep. This study addresses the following research gaps:
\begin{itemize}
    \item Haptic biofeedback has been minimally explored in sleep contexts, especially for sleep onset. Prior work has primarily focused on anxiety reduction under external stressors \cite{xu2021effect, costa_boostmeup_2019, t_azevedo_calming_2017}.
    \item Existing approaches often use music as background audio or as an auditory biofeedback channel \cite{yu2018unwind}; our system instead encodes musical meter structures directly into the haptic signal.
    \item Most wearable haptic systems rely on open-loop signals (e.g., fixed-frequency based heart rate modulation); in contrast, we implement a closed-loop system that adapts in real-time based on the participant's heart rates.
\end{itemize}

\section{Study 1: Relaxation Effects of Haptic Patterns}
\label{sec:study1}

In the first study, we tested five vibration pattern samples informed by lullaby characteristics and biofeedback mechanisms, comparing them to a non-vibration control condition. To support this, we developed a custom smartwatch application capable of generating rhythmic patterns and integrating heart rate sensing for dynamic biofeedback control. The effects of the patterns on relaxation and user experience were assessed through heart rate analysis, perceived arousal, vibration experience, and user preference.

\subsection{Methods}

\subsubsection{System}
\label{subsubsec:study1_system}
\begin{figure}[ht]
    \centering
    \includegraphics[width=0.8\linewidth]{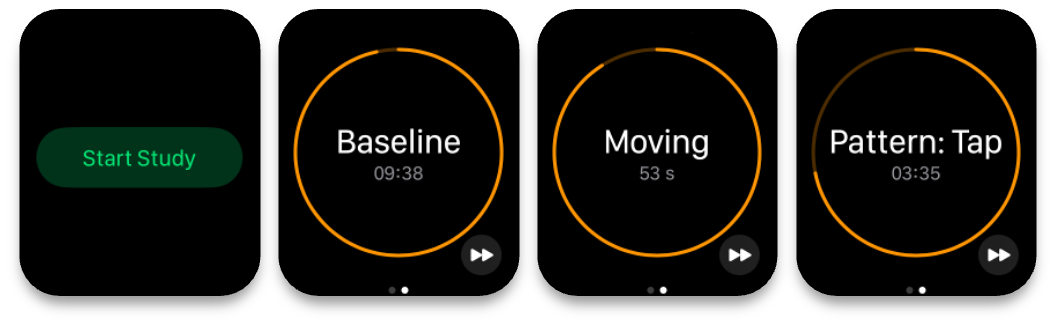}
\caption{Screenshots of the custom Apple Watch app used in Study~1 to guide the procedure, showing interface elements for study start, baseline heart rate recording, pre-condition movement, and vibration condition.}

    \label{fig:prestudy_app}
\end{figure}

We developed an Apple Watch application \cite{AppleWatchKitHaptic} (\autoref{fig:prestudy_app}) for customizing and playing vibrotactile feedback patterns in real-time biofeedback sessions. The app delivered vibration and recorded physiological data during Study~1. Vibrotactile feedback was generated using Apple’s WatchKit framework and the built-in Taptic Engine to play predefined patterns~\cite{apple_watchkit_play}. Real-time heart rate data was accessed via Apple HealthKit \cite{AppleHealthKitHeartRate} to enable adaptive stimulation. During sessions, the app retrieved the participant’s heart rate and dynamically adjusted vibration frequency based on current heart rate, aiming to support adaptive entrainment and personalized relaxation.
The application provided a fully guided workflow to standardize the experimental procedure. Participants were led through all study stages directly on the watch, including baseline acquisition, pattern selection, biofeedback-driven playback, and prompts for self-assessment questionnaires. This standalone interface enabled autonomous operation without researcher intervention or companion devices. All relevant data, including heart rate samples and event timestamps, were logged locally in CSV format. Upon session completion, the data were automatically transferred to the paired iPhone via the WatchConnectivity framework \cite{AppleWatchConnectivity}, allowing secure and efficient retrieval for analysis.

\subsubsection{Vibration Pattern Design}

\begin{figure}[!ht]
    \centering
    \includegraphics[width=\linewidth]{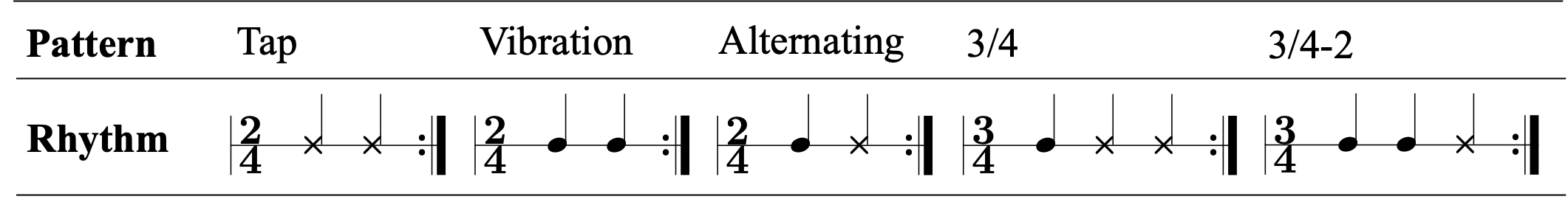}
\caption{Five different rhythmic patterns used in Study 1}
\label{fig:patterns}
\end{figure}

Five lullaby-inspired vibration rhythms were created and compared (\autoref{fig:patterns}). Each rhythm was constructed using two types of haptic elements as building blocks from the Apple Watch: a single tap ('Start' signal) and a short vibration ('Failure' signal). 
Lullabies commonly follow time signatures such as 4/4, 2/4, or 3/4 \cite{gurak2017book}. For instance, the 3/4 time signature, also known as the waltz rhythm, is associated with a gentle, swaying motion that conveys comfort and familiarity \cite{fernandez2016influence}.

The five rhythmic patterns were defined as follows  (\autoref{fig:patterns}):
\begin{itemize}
    \item Tab: A sequence of regular taps in a 2/4 time signature, representing accented beats.
    \item Vibration: A sequence of short vibrations in a 2/4 time signature, representing unaccented beats.
    \item Alternating: A 2/4 time signature alternating between taps and vibrations.
    \item Alternating 3/4: A 3/4 time signature alternating between taps and vibrations (referred to as \textit{3/4}).
    \item Alternating 3/4-2: A variation of the previous alternating 3/4 pattern, designed as a structural counterpart (referred to as \textit{3/4-2}).
\end{itemize}

\subsubsection{Biofeedback Adaptation}

To reflect the principle of biofeedback, which supports users in monitoring and modulating their own physiological signals, we implemented a minimal frequency adjustment based on real-time heart rate.
Evidence from a nap study suggests that heart rate biofeedback with a subtle reduction of approximately 3–5\% is effective for “weak non-invasive forcing” \cite{choi2019effect, anishchenko2000entrainment}, where the mean of recorded heart rate of 70.38 BPM, falls within the low-tempo range (60--80 BPM), aligning with tempo characteristics commonly found in music people choose for sleep \cite{lai2006music}.
Hence, we used the 4\% reduction of current heart rate to enforce physiological relaxation (configured every second). This approach contrasts with previous passive haptic biofeedback contexts, such as anxiety mitigation, where larger frequency adjustments (20–30\%) are often used while participants are engaged in cognitively demanding foreground tasks \cite{costa_boostmeup_2019, t_azevedo_calming_2017}, justifying a stronger entrainment strategy.

\subsubsection{Procedure}

\begin{figure*}[!ht]
    \centering
    \includegraphics[width=0.8\linewidth]{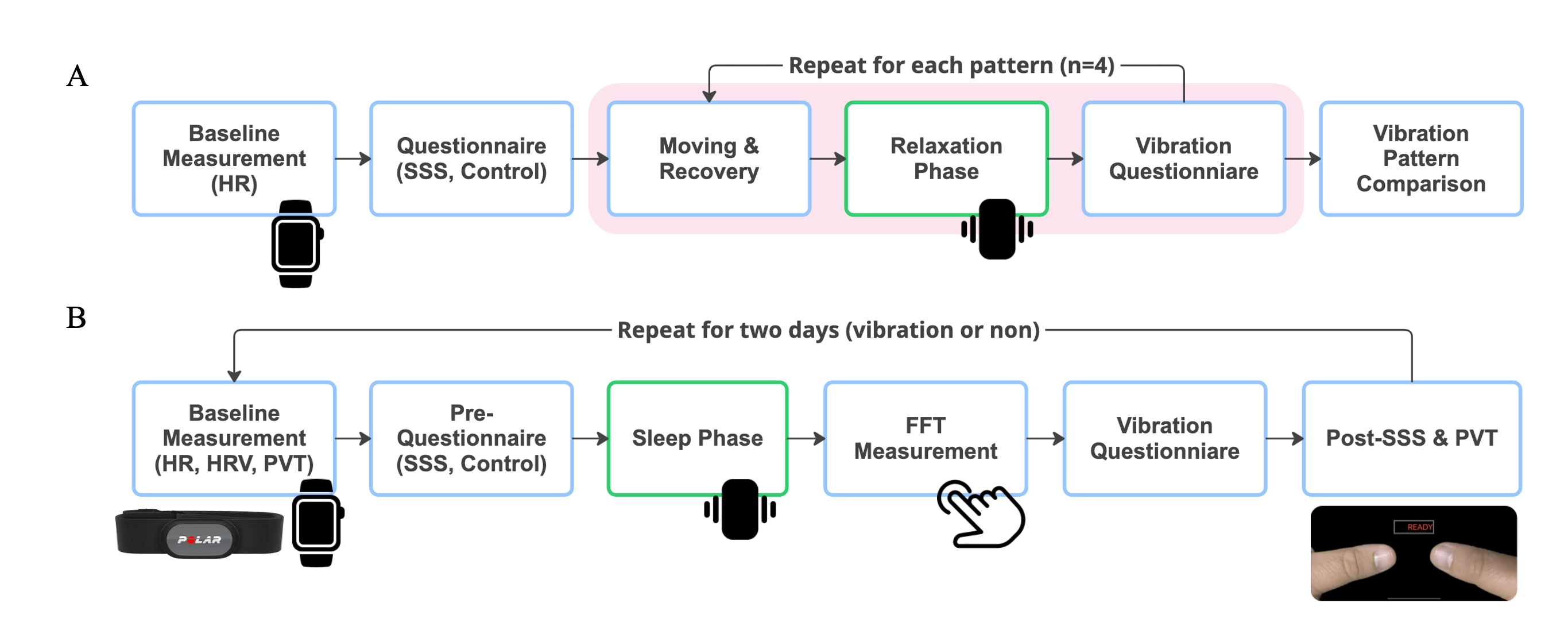}
    \caption{Overview of the procedures for (A) Study~1 and (B) Study~2}
    \label{fig:procedures_all}
\end{figure*}

In a within-subject design, a total of 20 healthy participants (15 males, 5 females; age $M{=}27.65$, $SD{=}10.69$) assessed five vibration patterns (\autoref{fig:patterns}). The study was conducted in a controlled environment, where participants sat at a table. Informed consent was obtained from all participants following a briefing on the study's procedures and objectives. The study procedure is outlined in \autoref{fig:procedures_all}. 

At the beginning of the study, participants wore an Apple Watch (SE 2, 40mm) and configured the custom app with a participant ID. Participants completed a ten-minute baseline using the smartwatch while engaging in light reading, with the final five minutes used for analysis to ensure heart rate stabilization \cite{speed2023measure}.
They then completed the Stanford Sleepiness Scale (SSS) for arousal assessment, ranging from "feeling active" or "wide awake" to "sleep onset soon" \cite{shahid2012stanford, sonclosed}, and reported recent tobacco and caffeine intake as control factors.
Participants subsequently experienced five vibration patterns in Latin square design to control for order effects. Before each condition, they walked slowly for one minute followed by a one-minute recovery to standardize pre-condition activity and reduce carryover effects.
Each pattern was presented for five minutes to balance study duration and response quality \cite{meterko2015response}. Heart rate was recorded throughout using the same smartwatch that delivered the vibrations \cite{shcherbina2017accuracy}. After each condition, participants completed a questionnaire evaluating the vibration's effects on relaxation (\emph{comfort}, \emph{relaxation}, \emph{sleepiness}) and ambience (\emph{recognizability}, \emph{choppiness}), using a five-point Likert scale.
After all exposures, participants ranked the patterns by relaxation suitability and provided open feedback. Snacks were offered as appreciation.

\subsection{Results}

We used a Friedman test to evaluate differences across multiple conditions, with Wilcoxon signed-rank tests as post-hoc pairwise comparisons. A significance level of $\alpha {=} 0.05$ was used for all statistical tests. 

The evaluation revealed consistent trends in physiological, perceptual, and preference measures \autoref{fig:study1_hrrank}. All vibration patterns significantly reduced heart rate compared to baseline, although their effects were largely comparable. While subjective arousal levels did not differ significantly across patterns, perceived relaxation favored the 3/4 and Alternating rhythms. Monotonous patterns, particularly those with identical consecutive patterns such as Tap and Vibration were perceived as less pleasant. Although not statistically significant, Tap and Vibration consistently received lower ratings, suggesting they were perceived as less comfortable and more choppy. Preference rankings further confirmed the positive reception of the 3/4 pattern, which emerged as the most favored among participants.

\subsubsection{Heart Rate}
\begin{figure}
    \centering
    \includegraphics[width=\linewidth]{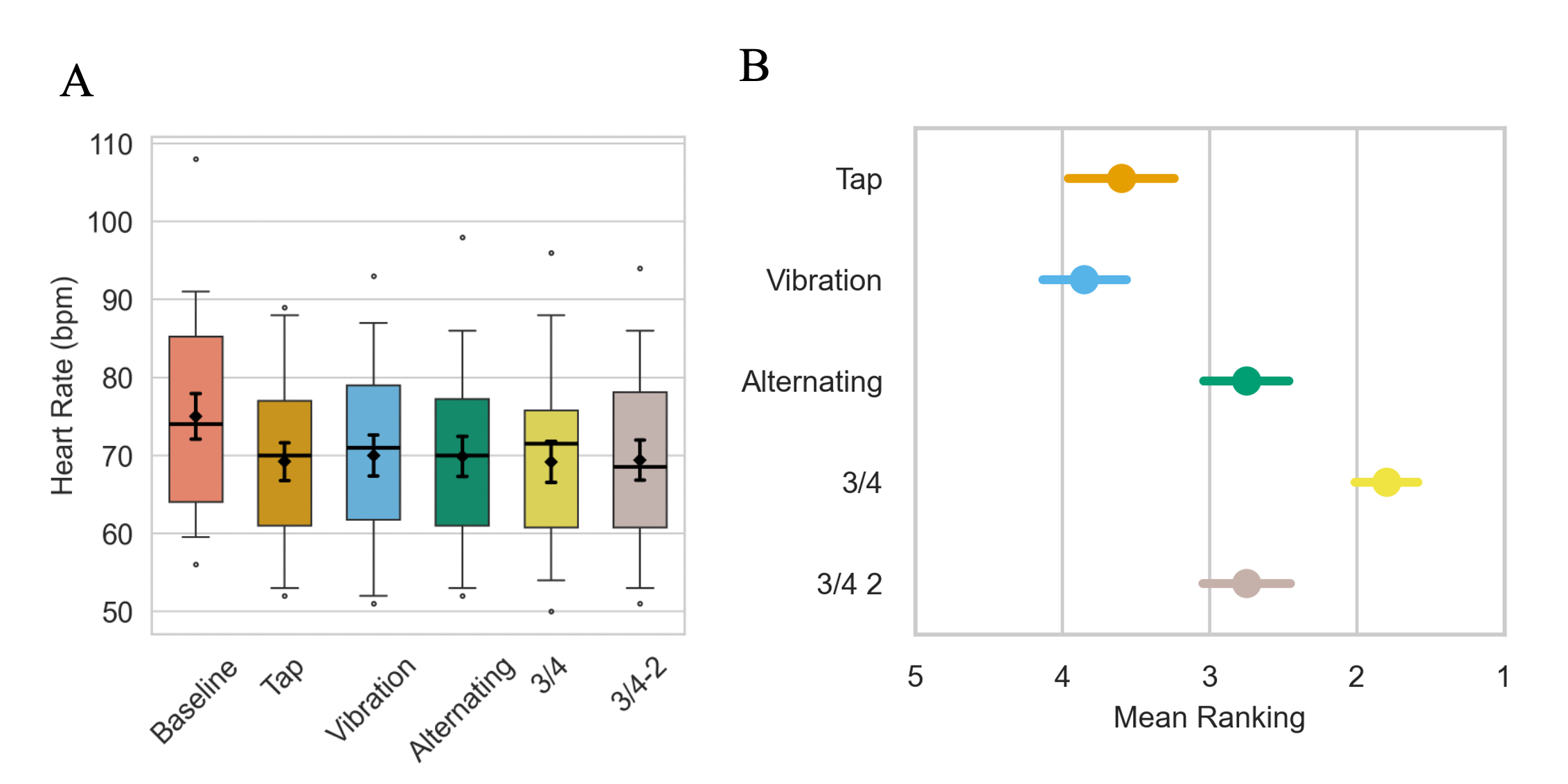}
    \caption{(A) Median heart rate across conditions. Overlaid point markers represent means with standard errors. (B) Preference rankings of vibration conditions (1 {=} most preferred).}

    \label{fig:study1_hrrank}
\end{figure}

A Friedman test on the median heart rate values revealed a significant difference between conditions ($\chi^2 {=} 27.172$, $p {<} .001$; \autoref{fig:study1_hrrank}). Post-hoc Wilcoxon signed-rank tests showed that all vibration patterns significantly reduced heart rate compared to the baseline: 3/4 ($p {<} .001$), 3/4-2 ($p {<} .001$), Alternating ($p {<} .01$), Tap ($p {<} .001$), and Vibration ($p {<} .01$). No significant differences were found between the vibration patterns themselves.

\subsubsection{Perceived Arousal}
Subjective arousal, assessed using the Stanford Sleepiness Scale (SSS), did not differ significantly between patterns according to a Friedman test ($p{=}.12$), with mean ratings ranging from 3.10 (Baseline) to 3.65 (Alternating), corresponding to feeling “awake, but relaxed; responsive but not fully alert”.

\subsubsection{Vibration Questionnaire}
\begin{figure}[!ht]
    \centering
    \includegraphics[width=
   0.9\linewidth]{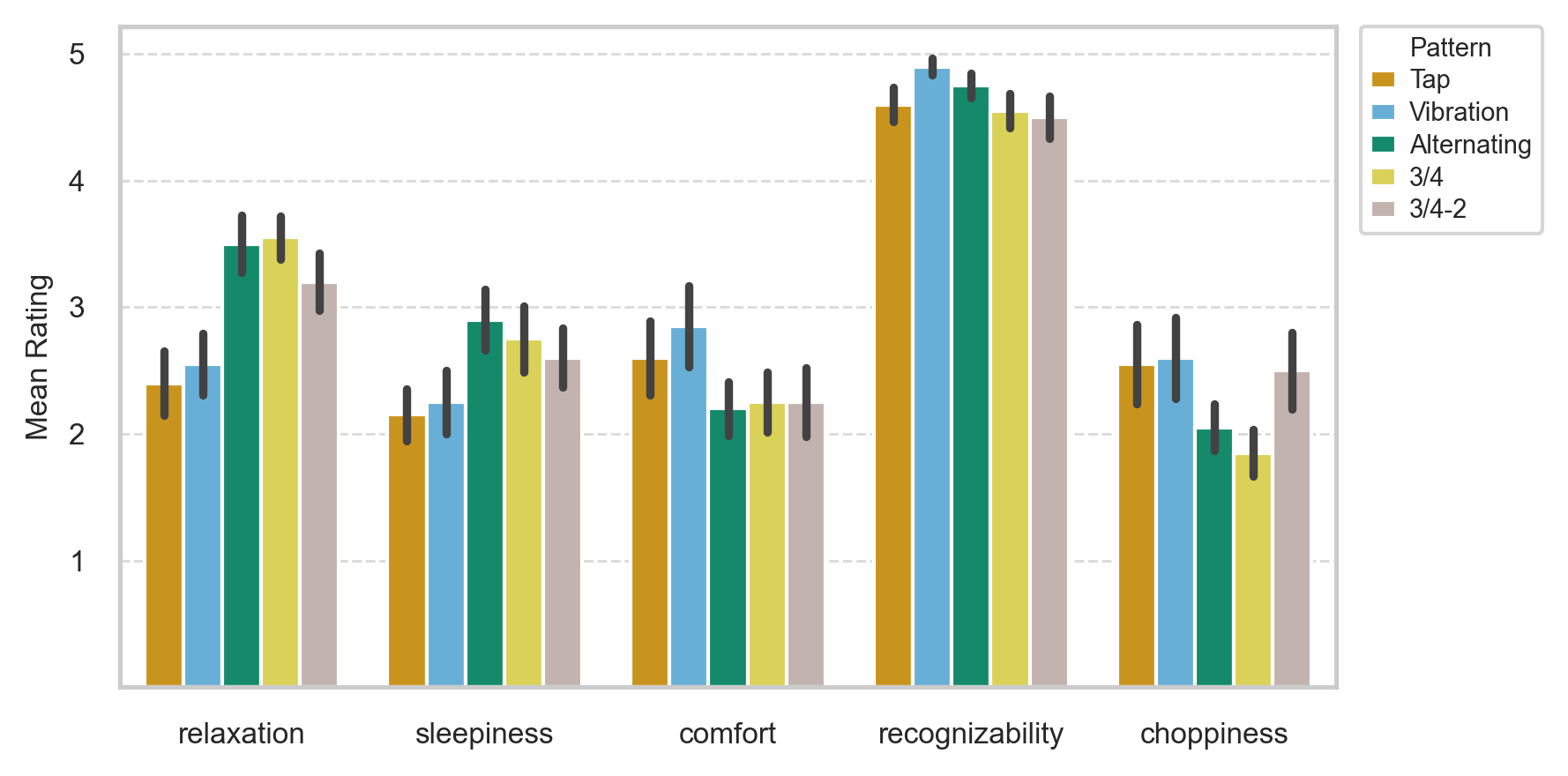}
    \caption{Mean ratings of vibration experiences across vibration patterns}
    \label{fig:study1_questionnaire}
\end{figure}

A Friedman test revealed a significant difference in perceived relaxation between patterns ($\chi^2{=}20.04$, $p{<}.001$; \autoref{fig:study1_questionnaire}). 
Post-hoc Wilcoxon signed-rank tests showed that 3/4 and Alternating ($M{=}3.55$, $M{=}3.50$) were rated significantly more relaxing than Tap ($M{=}2.40$; both $p{<}.05$) and Vibration ($M{=}2.55$; both $p{<}.05$).
No other items showed significant differences between patterns:  sleepiness ($\chi^2{=}7.88$, $p{=}.1$), comfort ($\chi^2{=}4.01$, $p{=}.40$), recognizability ($\chi^2{=}8.35$, $p{=}.08$), and choppiness ($\chi^2{=}5.92$, $p{=}.21$).

\subsubsection{Preference Ranking}
A Friedman test on the ranking data revealed a significant difference between patterns ($\chi^2{=}19.37$, $p{<}.01$). Post-hoc Wilcoxon tests showed that 3/4 was ranked significantly higher than all other patterns, including Tap and Vibration ($p{<}.01$, $p{<}.001$), as well as Alternating and 3/4-2 (both $p{<}.05$).
The 3/4 pattern was most frequently ranked first, selected by 50\% of participants, reinforcing its positive reception in the questionnaire evaluation. The 3/4-2 and Alternating patterns followed, each chosen as the top preference by 25\%.

\section{ Study 2: Single Pattern Effects on Relaxation and Sleep Onset}
\label{sec:main_study}

In the second study, we examined the prolonged effects of musically-informed vibrotactile biofeedback based on Study 1 on prolonged relaxation and sleep onset. Based on the results from Study 1, where the alternating 3/4 pattern effectively reduced heart rate and was positively perceived and most preferred for relaxation, we tested this pattern in a longer experimental session, during which participants attempted to sleep with or without vibration. Relaxation measures (heart rate, heart rate variability, perceived arousal) and sleep induction metrics (sleep onset, psychomotor performance) were analyzed.

\subsection{Methods}

\subsubsection{System}
Study~2 used an extended version of the Apple Watch application from Study 1 (Section \ref{subsubsec:study1_system}), which continued to deliver rhythmic biofeedback-based vibration and autonomously guided the procedure.
To support the protocol, a custom iPhone app was introduced. It administered the Finger Tapping Task (FTT), recorded tap events with precise timestamps, and established a Bluetooth Low Energy (BLE) connection to a Polar H9 chest strap for continuous electrocardiogram (ECG) recording. All physiological data were logged locally in CSV format for later analysis. To minimize disruption during sleep onset, the smartphone display remained off throughout the sleep phase, presenting no visual stimuli. A scheduled auditory alarm gently woke participants at the end of the session.
A Psychomotor Vigilance Test (PVT)~\cite{sonclosed} was conducted using the NASA-PVT iOS application~\cite{nasapvt_app} on an iPad Mini tablet, which recorded reaction times to visual stimuli through continuous screen-tapping input.

\subsubsection{Procedure}
A within-subject study was conducted with 28 healthy participants (18 males, 10 females; age $M{=}24.79$, $SD{=}6.30$) in a controlled environment furnished with a couch and a table with chairs. Each session included a 20-minute sleep phase and was scheduled between 4--8~PM to minimize disruption to participants’ natural sleep rhythms and ensure consistency across varying individual sleep schedules~\cite{cyr2022effect}. All participants completed both conditions (with and without vibrotactile stimulation) on two separate days to prevent carryover effects. The order of conditions was counterbalanced. The study procedure is summarized in \autoref{fig:procedures_all}.

Participants provided informed consent and were equipped with an Apple Watch and a Polar H9 chest strap~\cite{polar2024h9} to continuously record ECG data for heart rate and heart rate variability. After configuring the custom iPhone application, a 10-minute baseline period followed, during which participants engaged in light reading, as in Study~1.
They then completed a 5-minute baseline PVT \cite{dinges1985microcomputer}. Participants also filled out the Stanford Sleepiness Scale (SSS)~\cite{shahid2012stanford} to assess subjective arousal, and a control questionnaire covering sleep habits, stimulant use, physical activity, and meal timing.
In the sleep phase, participants lay on a couch with eyes closed for 20 minutes to relax and fall asleep, modeled after the Multiple Sleep Latency Test (MSLT)~\cite{carskadon1986guidelines}. During this period, the 3/4 vibrotactile pattern was delivered via the Apple Watch.
Following the sleep phase, participants completed a second PVT and post-study SSS to assess changes in alertness and subjective arousal. Snacks were offered as a token of appreciation upon completion.

\subsubsection{Measures and Analysis}
To evaluate the effects of rhythmic biofeedback vibration on relaxation and sleep onset, we collected physiological, behavioral, and subjective measures. In addition to the SSS ratings \cite{shahid2012stanford}, ECG data from the Polar H9 provided heart rate and heart rate variability, which were analyzed in both the time domain (RMSSD)~\cite{sztajzel2004heart, electrophysiology1996heart} and frequency domain—high-frequency (HF, 0.15--0.4\,Hz) and low-frequency (LF, 0.04--0.15\,Hz) components, and the LF/HF ratio~\cite{trinder2001autonomic}—to assess autonomic nervous system activity~\cite{mendes2009assessing, sztajzel2004heart, shinar2006autonomic}. 
HR and frequency-domain HRV were analyzed as change from baseline of the recorded day to control for individual differences, while RMSSD was directly compared due to its stability in absolute terms \cite{laborde2017heart, shaffer2017overview}.
SOL, defined as the time until the first tapping interval exceeds 8 seconds \cite{casagrande1997finger}, was derived from the FTT that recorded inter-tapping intervals (ITI) as behavioral indicators of sleepiness~\cite{curcio2001sleepiness, casagrande1997finger, ogilvie1984detection}. These measures have been shown to correlate with EEG-based markers distinguishing wakefulness from stage 1 sleep~\cite{ogilvie1984detection}. 
Alertness was assessed using the PVT \cite{dinges1985microcomputer}, with key metrics including reaction time, lapses (missed responses or responses $>$500\,ms), false starts, and errors~\cite{arsintescu2019validation}. Prior research has shown that drowsiness significantly slows reaction times and increases omission errors~\cite{dorrian2004psychomotor}.

\subsection{Results}
The same statistical approaches as in Study 1 were applied, including the Friedman test and Wilcoxon signed-rank tests, with a significance level of $\alpha {=} 0.05$. Of the initial 28 participants, three of them were excluded from the analysis due to unusually sudden high HR spikes observed during the sleep phase. 

Heart rate variability (HRV) measures, i.e., RMSSD and HF power, showed trends suggestive of increased parasympathetic activity during the vibration condition, while no statistically significant differences were observed between the vibration and control conditions in heart rate, HRV, perceived arousal, sleep onset latency, or psychomotor performance. Cognitive and behavioral outcomes remained consistent across conditions.

\subsubsection{Heart Rate and Heart Rate Variability}
A Wilcoxon signed-rank test on median heart rate differences (from baseline to sleep phase) between the 3/4 vibration condition and the control condition revealed no significant effect ($W{=}175.00$, $p{=}.75$). 
A main effect on RMSSD was not statistically significant, although the Wilcoxon signed-rank test indicated a trend toward increased parasympathetic activity with the 3/4 vibration pattern compared to baseline ($W{=}104$, $p{=}.07$). 
In the frequency domain analysis of HRV, no significant differences were found between conditions in the HF band ($W{=}128.00$, $p{=}.24$) or the LF/HF ratio ($W{=}141.00$, $p{=}.39$). While not statistically significant, both measures showed trends consistent with increased parasympathetic activity during the sleep phase, as indicated by a tendency toward higher HF power and a lower LF/HF ratio under vibration feedback.

\begin{figure}
    \centering
    \includegraphics[width=\linewidth]{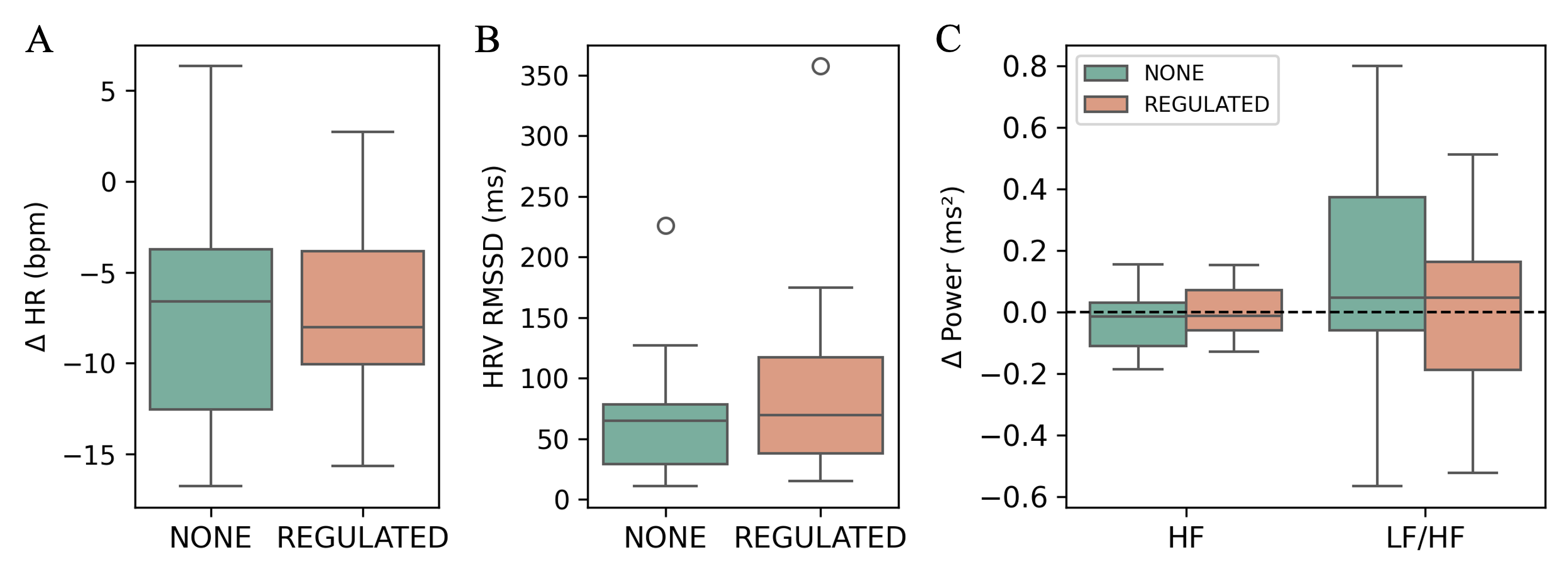}
    \caption{Physiological indicators of autonomic activity in vibration ("regulated") and control ("none") conditions: (A) Heart rate difference from baseline, (B) RMSSD (time-domain HRV), and (C) HF and LF/HF ratio (frequency-domain HRV) differences from baseline. Higher RMSSD and HF, and lower LF/HF values reflect increased parasympathetic activity.}

    \label{fig:study2_hrhrv}
\end{figure}

\subsubsection{Cognitive and Behavioral Outcomes}
No significant differences were observed between the 3/4 vibration and control conditions in cognitive or behavioral measures. Subjective arousal ratings on the SSS did not differ between conditions ($p{=}.13$). 
Sleep onset latency (SOL), derived from the FTT, showed no significant difference across conditions ($p{=}.82$), with no observable different across two days ($p{=}.46$).
Psychomotor performance, assessed using the PVT, revealed no significant differences in reaction time ($p{=}.15$), number of errors ($p{=}.41$), lapses ($p{=}.25$), or false starts ($p{=}.11$).

\section{Discussion}

The results indicate that biofeedback-based rhythmic vibration patterns effectively reduced heart rate and were perceived as relaxing, suggesting potential for acute physiological and subjective calming effects. However, when the most promising pattern was tested in a prolonged sleep initiation context, no statistically significant effects were observed across physiological, behavioral, or subjective measures. This discrepancy suggests that while rhythmic haptic feedback may support momentary relaxation, its influence on longer-term outcomes such as sleep onset may be limited or may require extended exposure, individual adaptation, or multimodal integration.

In Study~1, all vibration patterns led to heart rate reduction, but those with rhythmic variation---namely, 3/4 and Alternating---were rated more positively in terms of relaxation and user preference. In contrast, monotonous patterns such as Tap and Vibration were consistently rated as less relaxing, indicating that repetitive, unvaried feedback may be less effective for creating a pleasant or calming experience. Although no significant effects were found for comfort, it is noteworthy that alternating patterns (Alternating, 3/4, 3/4-2) were rated as more relaxing and sleep-inducing while also being perceived as less choppy. This suggests that simpler rhythms with shorter bursts may be experienced as more comfortable, but not necessarily more calming. High recognizability ratings across patterns may indicate a need to reduce perceptual salience in future designs. Unobtrusive feedback could help differentiate physiological responses and subjective arousal levels. Future studies may benefit from integrating minimally invasive sleep detection methods to further reduce interference with haptic feedback during sleep attempts.

In Study~2, the 3/4 vibration pattern did not yield statistically significant effects across measured outcomes, though trends in RMSSD and HF power---both indicators of vagal activity---suggested potential parasympathetic activation. While these effects were not robust enough to influence overt sleep initiation or behavioral alertness in a single session, they warrant further exploration in larger or longer-term studies to determine cumulative or individualized effects. Notably, the significant heart rate reduction observed in Study~1 may indicate that shorter stimulation durations (e.g., 5 minutes) are more effective, while extended exposure may attenuate the relaxation response or induce overstimulation. As one participant noted, "The vibration rhythm made me tired, but at a certain point, it could have stopped, as it seemed to keep me awake a little longer.'" This suggests that a 20-minute stimulation period might have been not ideal for sleep initiation, and future studies should explore shorter or adaptive durations.
In addition, some participants reported that the rhythmic feedback during the Finger Tapping Task (FTT) increased cognitive load, interfering with their ability to maintain a "personal rhythm" and potentially hindering sleep onset~\cite{goel2014cognitive}. This implies that dynamic patterns may conflict with internal rhythms and inadvertently induce cognitive stress when paired with tasks during sleep initiation.
Although not statistically significant, the vibration condition showed shorter reaction times (11.55\,ms vs. 18.70\,ms), fewer lapses, and fewer false starts compared to the control, suggesting a potential increase in alertness, possibly due to greater cognitive engagement.

\paragraph{Individual Differences and Moderating Factors.}
Some participants noted that the vibrations were distracting due to associations with smartwatch notifications or alarm functions. In Study~2, 19 participants reported not regularly using a smartwatch with haptic feedback. Among these non-regular users, vibration exposure was associated with a greater increase in HF power and a slight reduction in heart rate compared to baseline. Furthermore, the vibration condition appeared to mitigate declines in cognitive performance, with more pronounced effects observed in this subgroup. These observations suggest that habitual smartwatch use may influence the perceived and physiological effectiveness of haptic feedback. Future studies should consider smartwatch usage as a potential moderating factor, accounting for both habituation effects and the influence of wrist-based placement.
An exploratory analysis including caffeine consumption as a covariate in linear mixed-effects models suggested a non-significant trend toward reduced sleep duration when caffeine was consumed closer to bedtime ($p{=}0.18$). Future studies with larger samples may help clarify the impact of such confounding factors, accounting for individual variability in sleep and relaxation responses.
The First Night Effect (FNE), characterized by increased alertness and reduced sleep quality on the first night in unfamiliar environments~\cite{agnew1966first, toussaint1995first, tamaki2005examination}, may have contributed to the non-significant results in this study, despite the randomization of study nights. Future research could include a buffer day to mitigate FNE and ensure more stable baseline sleep behavior.

\paragraph{Extending the Dimensions of Wearable Haptic Interaction.}
Our findings partially align with those of Choi et al. \cite{choi2019effect}, who demonstrated significant HRV effects from closed-loop vibration feedback during full sleep cycles using a woofer embedded under a mattress. In contrast to our direct-contact smartwatch-based stimulation, their setup provided indirect vibrations, which may have been less intrusive. To reduce potential disturbance from direct skin contact, future work could personalize vibration intensity or explore alternative tactile modalities such as \emph{soft touch} actuation, which more closely resembles gentle human touch and may mitigate associations between vibrations and alert functions~\cite{choi2020ambienbeat, bontula_deep_2023, brown2021design, choi_aspire_2021, papadopoulou_affective_2019}. Building on Choi et al.'s design, future studies might also examine alternative body locations—such as the back or head—that are more closely associated with rest, potentially enhancing the effectiveness of haptic interventions.

\section{Future Work}

Based on the findings and study design insights, several directions emerge for future research. First, the absence of significant long-term effects despite acute heart rate reductions suggests the need to explore adaptive or time-limited stimulation protocols that optimize relaxation without risking overstimulation during extended exposure. Second, the influence of user familiarity with wearable haptics should be examined more systematically to account for habituation or alert associations, which may moderate the effectiveness of vibrotactile feedback.
Third, the study design could be refined by including a buffer day to mitigate the First Night Effect (FNE), thus ensuring more stable and representative baseline sleep behavior. Finally, future work should investigate alternative haptic modalities or body locations that align more naturally with sleep and rest (e.g., soft-touch feedback or stimulation at the back or head), potentially improving perceptual comfort.

\section{Conclusion}

This study examined rhythmic, biofeedback-driven vibrotactile patterns delivered via a smartwatch as a passive intervention for relaxation and sleep initiation. The first study demonstrated significant effects of brief stimulation on heart rate and perceived relaxation. The second study showed that heart rate variability trends indicated increased parasympathetic activity, although no significant effects were found in the prolonged sleep initiation context. This suggests that while rhythmic haptics may support momentary relaxation, their translation to sleep-specific outcomes requires further refinement.
Our method addresses a gap in prior research by introducing sleep-related, biofeedback-driven haptics using commercially available wearables, compared to existing auditory or open-loop approaches.
Future research should refine physiological and subjective markers to distinguish relaxation from sleepiness,  refine stimulation timing and modality, explore alternative haptic patterns and body sites, and use larger, longitudinal designs to better understand individual responsiveness and cumulative effects in real-world sleep hygiene applications.

\begin{acks}
This work is funded by the pilot program Core-Informatics of the Helmholtz Association (HGF) and by the Federal Ministry of Education and Research (BMBF) and the Baden-W{\"u}rttemberg Ministry of Science as part of the Excellence Strategy of the German Federal and State Governments.
\end{acks}

\bibliographystyle{ACM-Reference-Format}
\bibliography{references}


\begin{thebibliography}{73}


\ifx \showCODEN    \undefined \def \showCODEN     #1{\unskip}     \fi
\ifx \showISBNx    \undefined \def \showISBNx     #1{\unskip}     \fi
\ifx \showISBNxiii \undefined \def \showISBNxiii  #1{\unskip}     \fi
\ifx \showISSN     \undefined \def \showISSN      #1{\unskip}     \fi
\ifx \showLCCN     \undefined \def \showLCCN      #1{\unskip}     \fi
\ifx \shownote     \undefined \def \shownote      #1{#1}          \fi
\ifx \showarticletitle \undefined \def \showarticletitle #1{#1}   \fi
\ifx \showURL      \undefined \def \showURL       {\relax}        \fi
\providecommand\bibfield[2]{#2}
\providecommand\bibinfo[2]{#2}
\providecommand\natexlab[1]{#1}
\providecommand\showeprint[2][]{arXiv:#2}

\bibitem[Agnew~Jr et~al\mbox{.}(1966)]%
        {agnew1966first}
\bibfield{author}{\bibinfo{person}{HW Agnew~Jr}, \bibinfo{person}{Wilse~B Webb}, {and} \bibinfo{person}{Robert~L Williams}.} \bibinfo{year}{1966}\natexlab{}.
\newblock \showarticletitle{The first night effect: An Eeg studyof sleep}.
\newblock \bibinfo{journal}{\emph{Psychophysiology}} \bibinfo{volume}{2}, \bibinfo{number}{3} (\bibinfo{year}{1966}), \bibinfo{pages}{263--266}.
\newblock


\bibitem[Anishchenko et~al\mbox{.}(2000)]%
        {anishchenko2000entrainment}
\bibfield{author}{\bibinfo{person}{Vadim~S Anishchenko}, \bibinfo{person}{Alexander~G Balanov}, \bibinfo{person}{Natalia~B Janson}, \bibinfo{person}{Natalia~B Igosheva}, {and} \bibinfo{person}{Grigory~V Bordyugov}.} \bibinfo{year}{2000}\natexlab{}.
\newblock \showarticletitle{Entrainment between heart rate and weak noninvasive forcing}.
\newblock \bibinfo{journal}{\emph{International Journal of Bifurcation and Chaos}} \bibinfo{volume}{10}, \bibinfo{number}{10} (\bibinfo{year}{2000}), \bibinfo{pages}{2339--2348}.
\newblock


\bibitem[{Apple Inc.}(2015)]%
        {AppleWatchKitHaptic}
\bibfield{author}{\bibinfo{person}{{Apple Inc.}}} \bibinfo{year}{2015}\natexlab{}.
\newblock \bibinfo{title}{WKHapticType -- Apple Developer Dokumentation}.
\newblock \bibinfo{howpublished}{\url{https://developer.apple.com/documentation/watchkit/wkhaptictype}}.
\newblock
\newblock
\shownote{Zugriff am: 30.4.2024}.


\bibitem[{Apple Inc.}(2024a)]%
        {AppleHealthKitHeartRate}
\bibfield{author}{\bibinfo{person}{{Apple Inc.}}} \bibinfo{year}{2024}\natexlab{a}.
\newblock \bibinfo{title}{HKQuantityTypeIdentifier.heartRate}.
\newblock \bibinfo{howpublished}{\url{https://developer.apple.com/documentation/healthkit/hkquantitytypeidentifier/1615138-heartrate}}.
\newblock
\urldef\tempurl%
\url{https://developer.apple.com/documentation/healthkit/hkquantitytypeidentifier/1615138-heartrate}
\showURL{%
\tempurl}
\newblock
\shownote{Accessed: 2024-04-30}.


\bibitem[{Apple Inc.}(2024b)]%
        {apple_watchkit_play}
\bibfield{author}{\bibinfo{person}{{Apple Inc.}}} \bibinfo{year}{2024}\natexlab{b}.
\newblock \bibinfo{title}{WKInterfaceDevice play(\_:)- Method}.
\newblock
\urldef\tempurl%
\url{https://developer.apple.com/documentation/watchkit/wkinterfacedevice/play(_:)}
\showURL{%
\tempurl}
\newblock
\shownote{Accessed: 2024-04-30}.


\bibitem[Arsintescu et~al\mbox{.}(2019)]%
        {arsintescu2019validation}
\bibfield{author}{\bibinfo{person}{Lucia Arsintescu}, \bibinfo{person}{Kenji~H Kato}, \bibinfo{person}{Patrick~F Cravalho}, \bibinfo{person}{Nathan~H Feick}, \bibinfo{person}{Leland~S Stone}, {and} \bibinfo{person}{Erin~E Flynn-Evans}.} \bibinfo{year}{2019}\natexlab{}.
\newblock \showarticletitle{Validation of a touchscreen psychomotor vigilance task}.
\newblock \bibinfo{journal}{\emph{Accident Analysis \& Prevention}}  \bibinfo{volume}{126} (\bibinfo{year}{2019}), \bibinfo{pages}{173--176}.
\newblock


\bibitem[Association(2010)]%
        {american2010resolution}
\bibfield{author}{\bibinfo{person}{American~Medical Association}.} \bibinfo{year}{2010}\natexlab{}.
\newblock \showarticletitle{Resolution 503: Insufficient sleep in adolescents}.
\newblock \bibinfo{journal}{\emph{Chicago, IL: American Medical Association, American Academy of Sleep Medicine}} (\bibinfo{year}{2010}).
\newblock


\bibitem[Bernardi et~al\mbox{.}(2006)]%
        {bernardi2006cardiovascular}
\bibfield{author}{\bibinfo{person}{Luciano Bernardi}, \bibinfo{person}{Cesare Porta}, {and} \bibinfo{person}{Peter Sleight}.} \bibinfo{year}{2006}\natexlab{}.
\newblock \showarticletitle{Cardiovascular, cerebrovascular, and respiratory changes induced by different types of music in musicians and non-musicians: the importance of silence}.
\newblock \bibinfo{journal}{\emph{Heart}} \bibinfo{volume}{92}, \bibinfo{number}{4} (\bibinfo{year}{2006}), \bibinfo{pages}{445--452}.
\newblock


\bibitem[Bontula et~al\mbox{.}(2023)]%
        {bontula_deep_2023}
\bibfield{author}{\bibinfo{person}{Anisha Bontula}, \bibinfo{person}{Rhian~C. Preston}, \bibinfo{person}{Emily Shannon}, \bibinfo{person}{Cristina Wilson}, {and} \bibinfo{person}{Naomi~T. Fitter}.} \bibinfo{year}{2023}\natexlab{}.
\newblock \showarticletitle{Deep {Pressure} {Therapy}: {A} {Promising} {Anxiety} {Treatment} for {Individuals} with {High} {Touch} {Comfort}?}
\newblock \bibinfo{journal}{\emph{IEEE Transactions on Haptics}} (\bibinfo{year}{2023}).
\newblock
\urldef\tempurl%
\url{https://ieeexplore.ieee.org/abstract/document/10115454/}
\showURL{%
\tempurl}
\newblock
\shownote{Publisher: IEEE}.


\bibitem[Brown et~al\mbox{.}(2021)]%
        {brown2021design}
\bibfield{author}{\bibinfo{person}{Hannah Brown}, \bibinfo{person}{Emily Shannon}, {and} \bibinfo{person}{Naomi~T Fitter}.} \bibinfo{year}{2021}\natexlab{}.
\newblock \showarticletitle{Design and preliminary evaluation of the aid vest: An automatic inflatable wearable device for anxiety reduction}. In \bibinfo{booktitle}{\emph{2021 IEEE World Haptics Conference (WHC)}}. IEEE, \bibinfo{pages}{745--750}.
\newblock


\bibitem[Carskadon(1986)]%
        {carskadon1986guidelines}
\bibfield{author}{\bibinfo{person}{Mary~A Carskadon}.} \bibinfo{year}{1986}\natexlab{}.
\newblock \showarticletitle{Guidelines for the multiple sleep latency test (MSLT): a standard measure of sleepiness}.
\newblock \bibinfo{journal}{\emph{Sleep}} \bibinfo{volume}{9}, \bibinfo{number}{4} (\bibinfo{year}{1986}), \bibinfo{pages}{519--524}.
\newblock


\bibitem[Carskadon et~al\mbox{.}(2004)]%
        {carskadon2004regulation}
\bibfield{author}{\bibinfo{person}{Mary~A Carskadon}, \bibinfo{person}{Christine Acebo}, {and} \bibinfo{person}{Oskar~G Jenni}.} \bibinfo{year}{2004}\natexlab{}.
\newblock \showarticletitle{Regulation of adolescent sleep: implications for behavior}.
\newblock \bibinfo{journal}{\emph{Annals of the New York Academy of Sciences}} \bibinfo{volume}{1021}, \bibinfo{number}{1} (\bibinfo{year}{2004}), \bibinfo{pages}{276--291}.
\newblock


\bibitem[Casagrande et~al\mbox{.}(1997)]%
        {casagrande1997finger}
\bibfield{author}{\bibinfo{person}{Maria Casagrande}, \bibinfo{person}{Luigi De~Gennaro}, \bibinfo{person}{Cristiano Violani}, \bibinfo{person}{Paride Braibanti}, {and} \bibinfo{person}{Mario Bertini}.} \bibinfo{year}{1997}\natexlab{}.
\newblock \showarticletitle{A finger-tapping task and a reaction time task as behavioral measures of the transition from wakefulness to sleep: Which task interferes less with the sleep onset process?}
\newblock \bibinfo{journal}{\emph{Sleep}} \bibinfo{volume}{20}, \bibinfo{number}{4} (\bibinfo{year}{1997}), \bibinfo{pages}{301--312}.
\newblock


\bibitem[Chattu et~al\mbox{.}(2018)]%
        {chattu2018global}
\bibfield{author}{\bibinfo{person}{Vijay~Kumar Chattu}, \bibinfo{person}{Md~Dilshad Manzar}, \bibinfo{person}{Soosanna Kumary}, \bibinfo{person}{Deepa Burman}, \bibinfo{person}{David~Warren Spence}, {and} \bibinfo{person}{Seithikurippu~R Pandi-Perumal}.} \bibinfo{year}{2018}\natexlab{}.
\newblock \showarticletitle{The global problem of insufficient sleep and its serious public health implications}. In \bibinfo{booktitle}{\emph{Healthcare}}, Vol.~\bibinfo{volume}{7}. MDPI, \bibinfo{pages}{1}.
\newblock


\bibitem[Choi and Ishii(2020)]%
        {choi2020ambienbeat}
\bibfield{author}{\bibinfo{person}{Kyung~Yun Choi} {and} \bibinfo{person}{Hiroshi Ishii}.} \bibinfo{year}{2020}\natexlab{}.
\newblock \bibinfo{title}{ambienBeat: Wrist-worn Mobile Tactile Biofeedback for Heart Rate Rhythmic Regulation. TEI 2020 (2020), 17--30}.
\newblock


\bibitem[Choi et~al\mbox{.}(2021)]%
        {choi_aspire_2021}
\bibfield{author}{\bibinfo{person}{Kyung~Yun Choi}, \bibinfo{person}{Jinmo Lee}, \bibinfo{person}{Neska ElHaouij}, \bibinfo{person}{Rosalind Picard}, {and} \bibinfo{person}{Hiroshi Ishii}.} \bibinfo{year}{2021}\natexlab{}.
\newblock \showarticletitle{{aSpire}: {Clippable}, {Mobile} {Pneumatic}-{Haptic} {Device} for {Breathing} {Rate} {Regulation} via {Personalizable} {Tactile} {Feedback}}. In \bibinfo{booktitle}{\emph{Extended {Abstracts} of the 2021 {CHI} {Conference} on {Human} {Factors} in {Computing} {Systems}}} \emph{(\bibinfo{series}{{CHI} {EA} '21})}. \bibinfo{publisher}{Association for Computing Machinery}, \bibinfo{address}{New York, NY, USA}, \bibinfo{pages}{1--8}.
\newblock
\showISBNx{978-1-4503-8095-9}
\href{https://doi.org/10.1145/3411763.3451602}{doi:\nolinkurl{10.1145/3411763.3451602}}


\bibitem[Choi et~al\mbox{.}(2019)]%
        {choi2019effect}
\bibfield{author}{\bibinfo{person}{Sang~Ho Choi}, \bibinfo{person}{Heenam Yoon}, \bibinfo{person}{Hyung~Won Jin}, \bibinfo{person}{Hyun~Bin Kwon}, \bibinfo{person}{Seong~Min Oh}, \bibinfo{person}{Yu~Jin Lee}, {and} \bibinfo{person}{Kwang~Suk Park}.} \bibinfo{year}{2019}\natexlab{}.
\newblock \showarticletitle{Effect of closed-loop vibration stimulation on heart rhythm during naps}.
\newblock \bibinfo{journal}{\emph{Sensors}} \bibinfo{volume}{19}, \bibinfo{number}{19} (\bibinfo{year}{2019}), \bibinfo{pages}{4136}.
\newblock


\bibitem[Chua et~al\mbox{.}(2024)]%
        {chua2024know}
\bibfield{author}{\bibinfo{person}{Phoebe Chua}, \bibinfo{person}{Kat Agres}, {and} \bibinfo{person}{Suranga Nanayakkara}.} \bibinfo{year}{2024}\natexlab{}.
\newblock \showarticletitle{Know Thyself: Improving Interoceptive Ability through Ambient Biofeedback in the Workplace}. In \bibinfo{booktitle}{\emph{Proceedings of the 2023 Annual Workshop on Human-Computer Interaction (SIGHCI 2023)}}. \bibinfo{publisher}{Association for Information Systems (AIS)}, \bibinfo{address}{Hyderabad, India}.
\newblock
\newblock
\shownote{Available at: \url{https://aisel.aisnet.org/sighci2023/3}}.


\bibitem[Costa et~al\mbox{.}(2019)]%
        {costa_boostmeup_2019}
\bibfield{author}{\bibinfo{person}{Jean Costa}, \bibinfo{person}{François Guimbretière}, \bibinfo{person}{Malte~F. Jung}, {and} \bibinfo{person}{Tanzeem Choudhury}.} \bibinfo{year}{2019}\natexlab{}.
\newblock \showarticletitle{{BoostMeUp}: {Improving} {Cognitive} {Performance} in the {Moment} by {Unobtrusively} {Regulating} {Emotions} with a {Smartwatch}}.
\newblock \bibinfo{journal}{\emph{Proceedings of the ACM on Interactive, Mobile, Wearable and Ubiquitous Technologies}} \bibinfo{volume}{3}, \bibinfo{number}{2} (\bibinfo{date}{June} \bibinfo{year}{2019}), \bibinfo{pages}{40:1--40:23}.
\newblock
\href{https://doi.org/10.1145/3328911}{doi:\nolinkurl{10.1145/3328911}}


\bibitem[Curcio et~al\mbox{.}(2001)]%
        {curcio2001sleepiness}
\bibfield{author}{\bibinfo{person}{Giuseppe Curcio}, \bibinfo{person}{Maria Casagrande}, {and} \bibinfo{person}{Mario Bertini}.} \bibinfo{year}{2001}\natexlab{}.
\newblock \showarticletitle{Sleepiness: evaluating and quantifying methods}.
\newblock \bibinfo{journal}{\emph{International Journal of Psychophysiology}} \bibinfo{volume}{41}, \bibinfo{number}{3} (\bibinfo{year}{2001}), \bibinfo{pages}{251--263}.
\newblock


\bibitem[Cyr et~al\mbox{.}(2022)]%
        {cyr2022effect}
\bibfield{author}{\bibinfo{person}{Mari{\`e}ve Cyr}, \bibinfo{person}{Despina~Z Artenie}, \bibinfo{person}{Alain Al~Bikaii}, \bibinfo{person}{David Borsook}, {and} \bibinfo{person}{Jay~A Olson}.} \bibinfo{year}{2022}\natexlab{}.
\newblock \showarticletitle{The effect of evening light on circadian-related outcomes: A systematic review}.
\newblock \bibinfo{journal}{\emph{Sleep medicine reviews}}  \bibinfo{volume}{64} (\bibinfo{year}{2022}), \bibinfo{pages}{101660}.
\newblock


\bibitem[Dinges and Powell(1985)]%
        {dinges1985microcomputer}
\bibfield{author}{\bibinfo{person}{David~F Dinges} {and} \bibinfo{person}{John~W Powell}.} \bibinfo{year}{1985}\natexlab{}.
\newblock \showarticletitle{Microcomputer analyses of performance on a portable, simple visual RT task during sustained operations}.
\newblock \bibinfo{journal}{\emph{Behavior research methods, instruments, \& computers}} \bibinfo{volume}{17}, \bibinfo{number}{6} (\bibinfo{year}{1985}), \bibinfo{pages}{652--655}.
\newblock


\bibitem[Dobrushina et~al\mbox{.}(2024)]%
        {dobrushina2024training}
\bibfield{author}{\bibinfo{person}{Olga Dobrushina}, \bibinfo{person}{Yossi Tamim}, \bibinfo{person}{Iddo~Yehoshua Wald}, \bibinfo{person}{Amber Maimon}, {and} \bibinfo{person}{Amir Amedi}.} \bibinfo{year}{2024}\natexlab{}.
\newblock \showarticletitle{Training interoceptive awareness with real-time haptic vs. visual heartbeat feedback}.
\newblock \bibinfo{journal}{\emph{bioRxiv}} (\bibinfo{year}{2024}), \bibinfo{pages}{2024--01}.
\newblock


\bibitem[Dorrian et~al\mbox{.}(2004)]%
        {dorrian2004psychomotor}
\bibfield{author}{\bibinfo{person}{Jillian Dorrian}, \bibinfo{person}{Naomi~L Rogers}, {and} \bibinfo{person}{David~F Dinges}.} \bibinfo{year}{2004}\natexlab{}.
\newblock \showarticletitle{Psychomotor vigilance performance: Neurocognitive assay sensitive to sleep loss}.
\newblock \bibinfo{journal}{\emph{Sleep deprivation}} (\bibinfo{year}{2004}), \bibinfo{pages}{39--70}.
\newblock


\bibitem[Electro(2024)]%
        {polar2024h9}
\bibfield{author}{\bibinfo{person}{Polar Electro}.} \bibinfo{year}{2024}\natexlab{}.
\newblock \bibinfo{title}{Polar H9 Heart Rate Sensor}.
\newblock
\urldef\tempurl%
\url{https://www.polar.com/en/sensors/h9-heart-rate-sensor}
\showURL{%
\tempurl}
\newblock
\shownote{Accessed: 2024-04-07}.


\bibitem[Electrophysiology(1996)]%
        {electrophysiology1996heart}
\bibfield{author}{\bibinfo{person}{Task Force of the European Society of Cardiology the North American Society of~Pacing Electrophysiology}.} \bibinfo{year}{1996}\natexlab{}.
\newblock \showarticletitle{Heart rate variability: standards of measurement, physiological interpretation, and clinical use}.
\newblock \bibinfo{journal}{\emph{Circulation}} \bibinfo{volume}{93}, \bibinfo{number}{5} (\bibinfo{year}{1996}), \bibinfo{pages}{1043--1065}.
\newblock


\bibitem[Fern{\'a}ndez-Sotos et~al\mbox{.}(2016)]%
        {fernandez2016influence}
\bibfield{author}{\bibinfo{person}{Alicia Fern{\'a}ndez-Sotos}, \bibinfo{person}{Antonio Fern{\'a}ndez-Caballero}, {and} \bibinfo{person}{Jos{\'e}~M Latorre}.} \bibinfo{year}{2016}\natexlab{}.
\newblock \showarticletitle{Influence of tempo and rhythmic unit in musical emotion regulation}.
\newblock \bibinfo{journal}{\emph{Frontiers in computational neuroscience}}  \bibinfo{volume}{10} (\bibinfo{year}{2016}), \bibinfo{pages}{80}.
\newblock


\bibitem[Goel et~al\mbox{.}(2014)]%
        {goel2014cognitive}
\bibfield{author}{\bibinfo{person}{Namni Goel}, \bibinfo{person}{Takashi Abe}, \bibinfo{person}{Marcia~E Braun}, {and} \bibinfo{person}{David~F Dinges}.} \bibinfo{year}{2014}\natexlab{}.
\newblock \showarticletitle{Cognitive workload and sleep restriction interact to influence sleep homeostatic responses}.
\newblock \bibinfo{journal}{\emph{Sleep}} \bibinfo{volume}{37}, \bibinfo{number}{11} (\bibinfo{year}{2014}), \bibinfo{pages}{1745--1756}.
\newblock


\bibitem[Gurak(2017)]%
        {gurak2017book}
\bibfield{author}{\bibinfo{person}{Monika Gurak}.} \bibinfo{year}{2017}\natexlab{}.
\newblock \showarticletitle{The Book of Lullabies}.
\newblock  (\bibinfo{year}{2017}).
\newblock


\bibitem[Hertenstein et~al\mbox{.}(2009)]%
        {hertenstein2009communication}
\bibfield{author}{\bibinfo{person}{Matthew~J Hertenstein}, \bibinfo{person}{Rachel Holmes}, \bibinfo{person}{Margaret McCullough}, {and} \bibinfo{person}{Dacher Keltner}.} \bibinfo{year}{2009}\natexlab{}.
\newblock \showarticletitle{The communication of emotion via touch.}
\newblock \bibinfo{journal}{\emph{Emotion}} \bibinfo{volume}{9}, \bibinfo{number}{4} (\bibinfo{year}{2009}), \bibinfo{pages}{566}.
\newblock


\bibitem[Hesse(2013)]%
        {hesse2013musik}
\bibfield{author}{\bibinfo{person}{Horst-Peter Hesse}.} \bibinfo{year}{2013}\natexlab{}.
\newblock \bibinfo{booktitle}{\emph{Musik und Emotion: Wissenschaftliche Grundlagen des Musik-Erlebens}}.
\newblock \bibinfo{publisher}{Springer-Verlag}.
\newblock


\bibitem[Horv{\'a}th et~al\mbox{.}(2016)]%
        {horvath2016effects}
\bibfield{author}{\bibinfo{person}{Andr{\'a}s Horv{\'a}th}, \bibinfo{person}{Xavier Montana}, \bibinfo{person}{Jean-Pol Lanquart}, \bibinfo{person}{Philippe Hubain}, \bibinfo{person}{Anna Sz{\H{u}}cs}, \bibinfo{person}{Paul Linkowski}, {and} \bibinfo{person}{Gwenol{\'e} Loas}.} \bibinfo{year}{2016}\natexlab{}.
\newblock \showarticletitle{Effects of state and trait anxiety on sleep structure: A polysomnographic study in 1083 subjects}.
\newblock \bibinfo{journal}{\emph{Psychiatry research}}  \bibinfo{volume}{244} (\bibinfo{year}{2016}), \bibinfo{pages}{279--283}.
\newblock


\bibitem[Inc.({[n.\,d.]})]%
        {AppleWatchConnectivity}
\bibfield{author}{\bibinfo{person}{Apple Inc.}} \bibinfo{year}{[n.\,d.]}\natexlab{}.
\newblock \bibinfo{title}{WatchConnectivity Framework: Transferring Data Between iOS and watchOS Apps}.
\newblock \bibinfo{howpublished}{\url{https://developer.apple.com/documentation/watchconnectivity}}.
\newblock
\newblock
\shownote{Accessed: 2024-04-30}.


\bibitem[Jain et~al\mbox{.}(2020)]%
        {jain2020designing}
\bibfield{author}{\bibinfo{person}{Abhinandan Jain}, \bibinfo{person}{Adam~Haar Horowitz}, \bibinfo{person}{Felix Schoeller}, \bibinfo{person}{Sang-won Leigh}, \bibinfo{person}{Pattie Maes}, {and} \bibinfo{person}{Misha Sra}.} \bibinfo{year}{2020}\natexlab{}.
\newblock \showarticletitle{Designing interactions beyond conscious control: a new model for wearable interfaces}.
\newblock \bibinfo{journal}{\emph{Proceedings of the ACM on interactive, mobile, wearable and ubiquitous technologies}} \bibinfo{volume}{4}, \bibinfo{number}{3} (\bibinfo{year}{2020}), \bibinfo{pages}{1--23}.
\newblock


\bibitem[Jespersen(2022)]%
        {jespersen2022lullaby}
\bibfield{author}{\bibinfo{person}{Kira~Vibe Jespersen}.} \bibinfo{year}{2022}\natexlab{}.
\newblock \showarticletitle{A lullaby to the brain: The use of music as a sleep aid}.
\newblock In \bibinfo{booktitle}{\emph{The musical neurons}}. \bibinfo{publisher}{Springer}, \bibinfo{pages}{53--63}.
\newblock


\bibitem[Khalfa et~al\mbox{.}(2003)]%
        {khalfa2003effects}
\bibfield{author}{\bibinfo{person}{Stephanie Khalfa}, \bibinfo{person}{Simone~Dalla Bella}, \bibinfo{person}{Mathieu Roy}, \bibinfo{person}{Isabelle Peretz}, {and} \bibinfo{person}{Sonia~J Lupien}.} \bibinfo{year}{2003}\natexlab{}.
\newblock \showarticletitle{Effects of relaxing music on salivary cortisol level after psychological stress}.
\newblock \bibinfo{journal}{\emph{Annals of the New York Academy of Sciences}} \bibinfo{volume}{999}, \bibinfo{number}{1} (\bibinfo{year}{2003}), \bibinfo{pages}{374--376}.
\newblock


\bibitem[Knapp et~al\mbox{.}(1978)]%
        {knapp1978nonverbal}
\bibfield{author}{\bibinfo{person}{Mark~L Knapp}, \bibinfo{person}{Judith~A Hall}, {and} \bibinfo{person}{Terrence~G Horgan}.} \bibinfo{year}{1978}\natexlab{}.
\newblock \bibinfo{booktitle}{\emph{Nonverbal communication in human interaction}}. Vol.~\bibinfo{volume}{1}.
\newblock \bibinfo{publisher}{Holt, Rinehart and Winston New York}.
\newblock


\bibitem[Laborde et~al\mbox{.}(2017)]%
        {laborde2017heart}
\bibfield{author}{\bibinfo{person}{Sylvain Laborde}, \bibinfo{person}{Emma Mosley}, {and} \bibinfo{person}{Julian~F Thayer}.} \bibinfo{year}{2017}\natexlab{}.
\newblock \showarticletitle{Heart rate variability and cardiac vagal tone in psychophysiological research--recommendations for experiment planning, data analysis, and data reporting}.
\newblock \bibinfo{journal}{\emph{Frontiers in psychology}}  \bibinfo{volume}{8} (\bibinfo{year}{2017}), \bibinfo{pages}{213}.
\newblock


\bibitem[Lai and Good(2006)]%
        {lai2006music}
\bibfield{author}{\bibinfo{person}{Hui-Ling Lai} {and} \bibinfo{person}{Marion Good}.} \bibinfo{year}{2006}\natexlab{}.
\newblock \showarticletitle{Music improves sleep quality in older adults}.
\newblock \bibinfo{journal}{\emph{Journal of advanced nursing}} \bibinfo{volume}{53}, \bibinfo{number}{1} (\bibinfo{year}{2006}), \bibinfo{pages}{134--144}.
\newblock


\bibitem[Martire et~al\mbox{.}(2020)]%
        {martire2020stress}
\bibfield{author}{\bibinfo{person}{Viviana~Lo Martire}, \bibinfo{person}{Danila Caruso}, \bibinfo{person}{Laura Palagini}, \bibinfo{person}{Giovanna Zoccoli}, {and} \bibinfo{person}{Stefano Bastianini}.} \bibinfo{year}{2020}\natexlab{}.
\newblock \showarticletitle{Stress \& sleep: A relationship lasting a lifetime}.
\newblock \bibinfo{journal}{\emph{Neuroscience \& Biobehavioral Reviews}}  \bibinfo{volume}{117} (\bibinfo{year}{2020}), \bibinfo{pages}{65--77}.
\newblock


\bibitem[Mendes(2009)]%
        {mendes2009assessing}
\bibfield{author}{\bibinfo{person}{Wendy~Berry Mendes}.} \bibinfo{year}{2009}\natexlab{}.
\newblock \showarticletitle{Assessing autonomic nervous system activity}.
\newblock \bibinfo{journal}{\emph{Methods in social neuroscience}} \bibinfo{volume}{118}, \bibinfo{number}{147} (\bibinfo{year}{2009}), \bibinfo{pages}{21}.
\newblock


\bibitem[Meterko et~al\mbox{.}(2015)]%
        {meterko2015response}
\bibfield{author}{\bibinfo{person}{Mark Meterko}, \bibinfo{person}{Joseph~D Restuccia}, \bibinfo{person}{Kelly Stolzmann}, \bibinfo{person}{David Mohr}, \bibinfo{person}{Caitlin Brennan}, \bibinfo{person}{Justin Glasgow}, {and} \bibinfo{person}{Peter Kaboli}.} \bibinfo{year}{2015}\natexlab{}.
\newblock \showarticletitle{Response rates, nonresponse bias, and data quality: results from a national survey of senior healthcare leaders}.
\newblock \bibinfo{journal}{\emph{Public Opinion Quarterly}} \bibinfo{volume}{79}, \bibinfo{number}{1} (\bibinfo{year}{2015}), \bibinfo{pages}{130--144}.
\newblock


\bibitem[Morin and Espie(2011)]%
        {morin2011oxford}
\bibfield{author}{\bibinfo{person}{Charles~M Morin} {and} \bibinfo{person}{Colin~A Espie}.} \bibinfo{year}{2011}\natexlab{}.
\newblock \bibinfo{booktitle}{\emph{The Oxford handbook of sleep and sleep disorders}}.
\newblock \bibinfo{publisher}{Oxford University Press}.
\newblock


\bibitem[Murawski et~al\mbox{.}(2018)]%
        {murawski2018systematic}
\bibfield{author}{\bibinfo{person}{Beatrice Murawski}, \bibinfo{person}{Levi Wade}, \bibinfo{person}{Ronald~C Plotnikoff}, \bibinfo{person}{David~R Lubans}, {and} \bibinfo{person}{Mitch~J Duncan}.} \bibinfo{year}{2018}\natexlab{}.
\newblock \showarticletitle{A systematic review and meta-analysis of cognitive and behavioral interventions to improve sleep health in adults without sleep disorders}.
\newblock \bibinfo{journal}{\emph{Sleep medicine reviews}}  \bibinfo{volume}{40} (\bibinfo{year}{2018}), \bibinfo{pages}{160--169}.
\newblock


\bibitem[{NASA Ames Research Center}(2023)]%
        {nasapvt_app}
\bibfield{author}{\bibinfo{person}{{NASA Ames Research Center}}.} \bibinfo{year}{2023}\natexlab{}.
\newblock \bibinfo{title}{{NASA PVT [Mobile application software]}}.
\newblock \bibinfo{howpublished}{\url{https://apps.apple.com/app/id1146834402}}.
\newblock
\newblock
\shownote{Accessed: 2025-05-23}.


\bibitem[Ogilvie and Wilkinson(1984)]%
        {ogilvie1984detection}
\bibfield{author}{\bibinfo{person}{Robert~D Ogilvie} {and} \bibinfo{person}{Robert~T Wilkinson}.} \bibinfo{year}{1984}\natexlab{}.
\newblock \showarticletitle{The detection of sleep onset: behavioral and physiological convergence}.
\newblock \bibinfo{journal}{\emph{Psychophysiology}} \bibinfo{volume}{21}, \bibinfo{number}{5} (\bibinfo{year}{1984}), \bibinfo{pages}{510--520}.
\newblock


\bibitem[Ooishi et~al\mbox{.}(2017)]%
        {ooishi2017increase}
\bibfield{author}{\bibinfo{person}{Yuuki Ooishi}, \bibinfo{person}{Hideo Mukai}, \bibinfo{person}{Ken Watanabe}, \bibinfo{person}{Suguru Kawato}, {and} \bibinfo{person}{Makio Kashino}.} \bibinfo{year}{2017}\natexlab{}.
\newblock \showarticletitle{Increase in salivary oxytocin and decrease in salivary cortisol after listening to relaxing slow-tempo and exciting fast-tempo music}.
\newblock \bibinfo{journal}{\emph{PloS one}} \bibinfo{volume}{12}, \bibinfo{number}{12} (\bibinfo{year}{2017}), \bibinfo{pages}{e0189075}.
\newblock


\bibitem[Pagel and Parnes(2001)]%
        {pagel2001medications}
\bibfield{author}{\bibinfo{person}{JF Pagel} {and} \bibinfo{person}{Bennett~L Parnes}.} \bibinfo{year}{2001}\natexlab{}.
\newblock \showarticletitle{Medications for the treatment of sleep disorders: an overview}.
\newblock \bibinfo{journal}{\emph{Primary care companion to the Journal of clinical psychiatry}} \bibinfo{volume}{3}, \bibinfo{number}{3} (\bibinfo{year}{2001}), \bibinfo{pages}{118}.
\newblock


\bibitem[Palumbo et~al\mbox{.}(2017)]%
        {palumbo2017interpersonal}
\bibfield{author}{\bibinfo{person}{Richard~V Palumbo}, \bibinfo{person}{Marisa~E Marraccini}, \bibinfo{person}{Lisa~L Weyandt}, \bibinfo{person}{Oliver Wilder-Smith}, \bibinfo{person}{Heather~A McGee}, \bibinfo{person}{Siwei Liu}, {and} \bibinfo{person}{Matthew~S Goodwin}.} \bibinfo{year}{2017}\natexlab{}.
\newblock \showarticletitle{Interpersonal autonomic physiology: A systematic review of the literature}.
\newblock \bibinfo{journal}{\emph{Personality and social psychology review}} \bibinfo{volume}{21}, \bibinfo{number}{2} (\bibinfo{year}{2017}), \bibinfo{pages}{99--141}.
\newblock


\bibitem[Papadopoulou et~al\mbox{.}(2019)]%
        {papadopoulou_affective_2019}
\bibfield{author}{\bibinfo{person}{Athina Papadopoulou}, \bibinfo{person}{Jaclyn Berry}, \bibinfo{person}{Terry Knight}, {and} \bibinfo{person}{Rosalind Picard}.} \bibinfo{year}{2019}\natexlab{}.
\newblock \showarticletitle{Affective {Sleeve}: {Wearable} {Materials} with {Haptic} {Action} for {Promoting} {Calmness}}. In \bibinfo{booktitle}{\emph{Distributed, {Ambient} and {Pervasive} {Interactions}}}, \bibfield{editor}{\bibinfo{person}{Norbert Streitz} {and} \bibinfo{person}{Shin’ichi Konomi}} (Eds.). \bibinfo{publisher}{Springer International Publishing}, \bibinfo{address}{Cham}, \bibinfo{pages}{304--319}.
\newblock
\showISBNx{978-3-030-21935-2}
\href{https://doi.org/10.1007/978-3-030-21935-2_23}{doi:\nolinkurl{10.1007/978-3-030-21935-2_23}}


\bibitem[Sateia et~al\mbox{.}(2000)]%
        {sateia2000evaluation}
\bibfield{author}{\bibinfo{person}{Michael~J Sateia}, \bibinfo{person}{Karl Doghramji}, \bibinfo{person}{Peter~J Hauri}, {and} \bibinfo{person}{Charles~M Morin}.} \bibinfo{year}{2000}\natexlab{}.
\newblock \showarticletitle{Evaluation of chronic insomnia. An American Academy of Sleep Medicine review.}
\newblock \bibinfo{journal}{\emph{Sleep}} \bibinfo{volume}{23}, \bibinfo{number}{2} (\bibinfo{year}{2000}), \bibinfo{pages}{243--308}.
\newblock


\bibitem[Scarratt et~al\mbox{.}(2023)]%
        {scarratt2023audio}
\bibfield{author}{\bibinfo{person}{Rebecca~Jane Scarratt}, \bibinfo{person}{Ole~Adrian Heggli}, \bibinfo{person}{Peter Vuust}, {and} \bibinfo{person}{Kira~Vibe Jespersen}.} \bibinfo{year}{2023}\natexlab{}.
\newblock \showarticletitle{The audio features of sleep music: Universal and subgroup characteristics}.
\newblock \bibinfo{journal}{\emph{PloS One}} \bibinfo{volume}{18}, \bibinfo{number}{1} (\bibinfo{year}{2023}), \bibinfo{pages}{e0278813}.
\newblock


\bibitem[Shaffer and Ginsberg(2017)]%
        {shaffer2017overview}
\bibfield{author}{\bibinfo{person}{Fred Shaffer} {and} \bibinfo{person}{Jay~P Ginsberg}.} \bibinfo{year}{2017}\natexlab{}.
\newblock \showarticletitle{An overview of heart rate variability metrics and norms}.
\newblock \bibinfo{journal}{\emph{Frontiers in public health}}  \bibinfo{volume}{5} (\bibinfo{year}{2017}), \bibinfo{pages}{258}.
\newblock


\bibitem[Shahid et~al\mbox{.}(2012)]%
        {shahid2012stanford}
\bibfield{author}{\bibinfo{person}{Azmeh Shahid}, \bibinfo{person}{Kate Wilkinson}, \bibinfo{person}{Shai Marcu}, {and} \bibinfo{person}{Colin~M Shapiro}.} \bibinfo{year}{2012}\natexlab{}.
\newblock \showarticletitle{Stanford sleepiness scale (SSS)}.
\newblock \bibinfo{journal}{\emph{STOP, THAT and one hundred other sleep scales}} (\bibinfo{year}{2012}), \bibinfo{pages}{369--370}.
\newblock


\bibitem[Shcherbina et~al\mbox{.}(2017)]%
        {shcherbina2017accuracy}
\bibfield{author}{\bibinfo{person}{Anna Shcherbina}, \bibinfo{person}{C~Mikael Mattsson}, \bibinfo{person}{Daryl Waggott}, \bibinfo{person}{Heidi Salisbury}, \bibinfo{person}{Jeffrey~W Christle}, \bibinfo{person}{Trevor Hastie}, \bibinfo{person}{Matthew~T Wheeler}, {and} \bibinfo{person}{Euan~A Ashley}.} \bibinfo{year}{2017}\natexlab{}.
\newblock \showarticletitle{Accuracy in wrist-worn, sensor-based measurements of heart rate and energy expenditure in a diverse cohort}.
\newblock \bibinfo{journal}{\emph{Journal of personalized medicine}} \bibinfo{volume}{7}, \bibinfo{number}{2} (\bibinfo{year}{2017}), \bibinfo{pages}{3}.
\newblock


\bibitem[Shinar et~al\mbox{.}(2006)]%
        {shinar2006autonomic}
\bibfield{author}{\bibinfo{person}{Zvi Shinar}, \bibinfo{person}{Solange Akselrod}, \bibinfo{person}{Yaron Dagan}, {and} \bibinfo{person}{Armanda Baharav}.} \bibinfo{year}{2006}\natexlab{}.
\newblock \showarticletitle{Autonomic changes during wake--sleep transition: A heart rate variability based approach}.
\newblock \bibinfo{journal}{\emph{Autonomic Neuroscience}} \bibinfo{volume}{130}, \bibinfo{number}{1-2} (\bibinfo{year}{2006}), \bibinfo{pages}{17--27}.
\newblock


\bibitem[Slovak et~al\mbox{.}(2023)]%
        {slovak2023designing}
\bibfield{author}{\bibinfo{person}{Petr Slovak}, \bibinfo{person}{Alissa Antle}, \bibinfo{person}{Nikki Theofanopoulou}, \bibinfo{person}{Claudia Daud{\'e}n~Roquet}, \bibinfo{person}{James Gross}, {and} \bibinfo{person}{Katherine Isbister}.} \bibinfo{year}{2023}\natexlab{}.
\newblock \showarticletitle{Designing for emotion regulation interventions: an agenda for HCI theory and research}.
\newblock \bibinfo{journal}{\emph{ACM Transactions on Computer-Human Interaction}} \bibinfo{volume}{30}, \bibinfo{number}{1} (\bibinfo{year}{2023}), \bibinfo{pages}{1--51}.
\newblock


\bibitem[Son et~al\mbox{.}(2025)]%
        {sonclosed}
\bibfield{author}{\bibinfo{person}{Hyo~Won Son}, \bibinfo{person}{Sang~Hyuk Kim}, \bibinfo{person}{Tae~Mu Lee}, \bibinfo{person}{Hyoyoung Heo}, \bibinfo{person}{Hyun~Bin Kwon}, \bibinfo{person}{Byunghun Choi}, \bibinfo{person}{Heenam Yoon}, {and} \bibinfo{person}{Hyun~Jae Baek}.} \bibinfo{year}{2025}\natexlab{}.
\newblock \showarticletitle{Closed-loop vibration stimulation based on heart rhythm for reducing sleep inertia}.
\newblock \bibinfo{journal}{\emph{Journal of Sleep Research}} (\bibinfo{year}{2025}), \bibinfo{pages}{e14458}.
\newblock


\bibitem[Speed et~al\mbox{.}(2023)]%
        {speed2023measure}
\bibfield{author}{\bibinfo{person}{Cathy Speed}, \bibinfo{person}{Thomas Arneil}, \bibinfo{person}{Robert Harle}, \bibinfo{person}{Alex Wilson}, \bibinfo{person}{Alan Karthikesalingam}, \bibinfo{person}{Michael McConnell}, {and} \bibinfo{person}{Justin Phillips}.} \bibinfo{year}{2023}\natexlab{}.
\newblock \showarticletitle{Measure by measure: Resting heart rate across the 24-hour cycle}.
\newblock \bibinfo{journal}{\emph{PLOS Digital Health}} \bibinfo{volume}{2}, \bibinfo{number}{4} (\bibinfo{year}{2023}), \bibinfo{pages}{e0000236}.
\newblock


\bibitem[Sztajzel(2004)]%
        {sztajzel2004heart}
\bibfield{author}{\bibinfo{person}{Juan Sztajzel}.} \bibinfo{year}{2004}\natexlab{}.
\newblock \showarticletitle{Heart rate variability: a noninvasive electrocardiographic method to measure the autonomic nervous system}.
\newblock \bibinfo{journal}{\emph{Swiss medical weekly}} \bibinfo{volume}{134}, \bibinfo{number}{3536} (\bibinfo{year}{2004}), \bibinfo{pages}{514--522}.
\newblock


\bibitem[T.~Azevedo et~al\mbox{.}(2017)]%
        {t_azevedo_calming_2017}
\bibfield{author}{\bibinfo{person}{Ruben T.~Azevedo}, \bibinfo{person}{Nell Bennett}, \bibinfo{person}{Andreas Bilicki}, \bibinfo{person}{Jack Hooper}, \bibinfo{person}{Fotini Markopoulou}, {and} \bibinfo{person}{Manos Tsakiris}.} \bibinfo{year}{2017}\natexlab{}.
\newblock \showarticletitle{The calming effect of a new wearable device during the anticipation of public speech}.
\newblock \bibinfo{journal}{\emph{Scientific reports}} \bibinfo{volume}{7}, \bibinfo{number}{1} (\bibinfo{year}{2017}), \bibinfo{pages}{2285}.
\newblock
\urldef\tempurl%
\url{https://www.nature.com/articles/s41598-017-02274-2}
\showURL{%
\tempurl}
\newblock
\shownote{Publisher: Nature Publishing Group UK London}.


\bibitem[Tamaki et~al\mbox{.}(2005)]%
        {tamaki2005examination}
\bibfield{author}{\bibinfo{person}{Masako Tamaki}, \bibinfo{person}{Hiroshi Nittono}, \bibinfo{person}{Mitsuo Hayashi}, {and} \bibinfo{person}{Tadao Hori}.} \bibinfo{year}{2005}\natexlab{}.
\newblock \showarticletitle{Examination of the first-night effect during the sleep-onset period}.
\newblock \bibinfo{journal}{\emph{Sleep}} \bibinfo{volume}{28}, \bibinfo{number}{2} (\bibinfo{year}{2005}), \bibinfo{pages}{195--202}.
\newblock


\bibitem[Toussaint et~al\mbox{.}(1995)]%
        {toussaint1995first}
\bibfield{author}{\bibinfo{person}{Michel Toussaint}, \bibinfo{person}{Remy Luthringer}, \bibinfo{person}{Nicolas Schaltenbrand}, \bibinfo{person}{Gabriella Carelli}, \bibinfo{person}{Eric Lainey}, \bibinfo{person}{Anne Jacqmin}, \bibinfo{person}{Alain Muzet}, {and} \bibinfo{person}{Jean-Paul Macher}.} \bibinfo{year}{1995}\natexlab{}.
\newblock \showarticletitle{First-night effect in normal subjects and psychiatric inpatients}.
\newblock \bibinfo{journal}{\emph{Sleep}} \bibinfo{volume}{18}, \bibinfo{number}{6} (\bibinfo{year}{1995}), \bibinfo{pages}{463--469}.
\newblock


\bibitem[Trehub et~al\mbox{.}(1993)]%
        {trehub1993maternal}
\bibfield{author}{\bibinfo{person}{Sandra~E Trehub}, \bibinfo{person}{Anna~M Unyk}, {and} \bibinfo{person}{Laurel~J Trainor}.} \bibinfo{year}{1993}\natexlab{}.
\newblock \showarticletitle{Maternal singing in cross-cultural perspective}.
\newblock \bibinfo{journal}{\emph{Infant behavior and development}} \bibinfo{volume}{16}, \bibinfo{number}{3} (\bibinfo{year}{1993}), \bibinfo{pages}{285--295}.
\newblock


\bibitem[Trinder et~al\mbox{.}(2001)]%
        {trinder2001autonomic}
\bibfield{author}{\bibinfo{person}{John Trinder}, \bibinfo{person}{Jan Kleiman}, \bibinfo{person}{Melinda Carrington}, \bibinfo{person}{Simon Smith}, \bibinfo{person}{Sibilah Breen}, \bibinfo{person}{Nellie Tan}, {and} \bibinfo{person}{Young Kim}.} \bibinfo{year}{2001}\natexlab{}.
\newblock \showarticletitle{Autonomic activity during human sleep as a function of time and sleep stage}.
\newblock \bibinfo{journal}{\emph{Journal of sleep research}} \bibinfo{volume}{10}, \bibinfo{number}{4} (\bibinfo{year}{2001}), \bibinfo{pages}{253--264}.
\newblock


\bibitem[Unyk et~al\mbox{.}(1992)]%
        {unyk1992lullabies}
\bibfield{author}{\bibinfo{person}{Anna~M Unyk}, \bibinfo{person}{Sandra~E Trehub}, \bibinfo{person}{Laurel~J Trainor}, {and} \bibinfo{person}{E~Glenn Schellenberg}.} \bibinfo{year}{1992}\natexlab{}.
\newblock \showarticletitle{Lullabies and simplicity: A cross-cultural perspective}.
\newblock \bibinfo{journal}{\emph{Psychology of music}} \bibinfo{volume}{20}, \bibinfo{number}{1} (\bibinfo{year}{1992}), \bibinfo{pages}{15--28}.
\newblock


\bibitem[Van~Erp and Toet(2015)]%
        {van2015social}
\bibfield{author}{\bibinfo{person}{Jan~BF Van~Erp} {and} \bibinfo{person}{Alexander Toet}.} \bibinfo{year}{2015}\natexlab{}.
\newblock \showarticletitle{Social touch in human--computer interaction}.
\newblock \bibinfo{journal}{\emph{Frontiers in digital humanities}}  \bibinfo{volume}{2} (\bibinfo{year}{2015}), \bibinfo{pages}{2}.
\newblock


\bibitem[Vyas et~al\mbox{.}(2023)]%
        {vyas2023descriptive}
\bibfield{author}{\bibinfo{person}{Preeti Vyas}, \bibinfo{person}{Unma~Mayur Desai}, \bibinfo{person}{Karin Yamakawa}, {and} \bibinfo{person}{Karon Maclean}.} \bibinfo{year}{2023}\natexlab{}.
\newblock \showarticletitle{A Descriptive Analysis of a Formative Decade of Research in Affective Haptic System Design}. In \bibinfo{booktitle}{\emph{Proceedings of the 2023 CHI Conference on Human Factors in Computing Systems}}. \bibinfo{pages}{1--23}.
\newblock


\bibitem[Woodward et~al\mbox{.}(2020)]%
        {woodward2020beyond}
\bibfield{author}{\bibinfo{person}{Kieran Woodward}, \bibinfo{person}{Eiman Kanjo}, \bibinfo{person}{David~J Brown}, \bibinfo{person}{T~Martin McGinnity}, \bibinfo{person}{Becky Inkster}, \bibinfo{person}{Donald~J Macintyre}, {and} \bibinfo{person}{Athanasios Tsanas}.} \bibinfo{year}{2020}\natexlab{}.
\newblock \showarticletitle{Beyond mobile apps: a survey of technologies for mental well-being}.
\newblock \bibinfo{journal}{\emph{IEEE Transactions on Affective Computing}} \bibinfo{volume}{13}, \bibinfo{number}{3} (\bibinfo{year}{2020}), \bibinfo{pages}{1216--1235}.
\newblock


\bibitem[Xu et~al\mbox{.}(2021)]%
        {xu2021effect}
\bibfield{author}{\bibinfo{person}{Mingdi Xu}, \bibinfo{person}{Takeshi Tachibana}, \bibinfo{person}{Nana Suzuki}, \bibinfo{person}{Eiichi Hoshino}, \bibinfo{person}{Yuri Terasawa}, \bibinfo{person}{Norihisa Miki}, {and} \bibinfo{person}{Yasuyo Minagawa}.} \bibinfo{year}{2021}\natexlab{}.
\newblock \showarticletitle{The effect of haptic stimulation simulating heartbeats on the regulation of physiological responses and prosocial behavior under stress: The influence of interoceptive accuracy}.
\newblock \bibinfo{journal}{\emph{Biological Psychology}}  \bibinfo{volume}{164} (\bibinfo{year}{2021}), \bibinfo{pages}{108172}.
\newblock


\bibitem[Yamasato et~al\mbox{.}(2019)]%
        {yamasato2019characteristics}
\bibfield{author}{\bibinfo{person}{Ami Yamasato}, \bibinfo{person}{Mayu Kondo}, \bibinfo{person}{Shunya Hoshino}, \bibinfo{person}{Jun Kikuchi}, \bibinfo{person}{Shigeki Okino}, {and} \bibinfo{person}{Kenji Yamamoto}.} \bibinfo{year}{2019}\natexlab{}.
\newblock \showarticletitle{Characteristics of music to improve the quality of sleep}.
\newblock \bibinfo{journal}{\emph{Music and Medicine}} \bibinfo{volume}{11}, \bibinfo{number}{3} (\bibinfo{year}{2019}), \bibinfo{pages}{195--202}.
\newblock


\bibitem[Yu et~al\mbox{.}(2018)]%
        {yu2018unwind}
\bibfield{author}{\bibinfo{person}{Bin Yu}, \bibinfo{person}{Mathias Funk}, \bibinfo{person}{Jun Hu}, {and} \bibinfo{person}{Loe Feijs}.} \bibinfo{year}{2018}\natexlab{}.
\newblock \showarticletitle{Unwind: a musical biofeedback for relaxation assistance}.
\newblock \bibinfo{journal}{\emph{Behaviour \& Information Technology}} \bibinfo{volume}{37}, \bibinfo{number}{8} (\bibinfo{year}{2018}), \bibinfo{pages}{800--814}.
\newblock


\bibitem[Zhao et~al\mbox{.}(2023)]%
        {zhao2023affective}
\bibfield{author}{\bibinfo{person}{Yiran Zhao}, \bibinfo{person}{Yujie Tao}, \bibinfo{person}{Grace Le}, \bibinfo{person}{Rui Maki}, \bibinfo{person}{Alexander Adams}, \bibinfo{person}{Pedro Lopes}, {and} \bibinfo{person}{Tanzeem Choudhury}.} \bibinfo{year}{2023}\natexlab{}.
\newblock \showarticletitle{Affective Touch as Immediate and Passive Wearable Intervention}.
\newblock \bibinfo{journal}{\emph{Proceedings of the ACM on Interactive, Mobile, Wearable and Ubiquitous Technologies}} \bibinfo{volume}{6}, \bibinfo{number}{4} (\bibinfo{year}{2023}), \bibinfo{pages}{1--23}.
\newblock


\end{thebibliography}

\appendix

\end{document}